\documentclass[submission,copyright,creativecommons]{eptcs}
\providecommand{\event}{FOCLASA 2010} 

\usepackage{amsmath,bbm}
\usepackage{amssymb}
\usepackage{amsfonts,verbatim}
\usepackage{graphicx,lscape,color}

\usepackage{prooftree}

\newcommand{\bpel}{\textsc{BPEL}}
\newcommand{\chr}{\textsc{CHR}}
\newcommand{\creole}{\textsc{Creole}}
\newcommand{\datalog}{\textsc{Datalog}}
\newcommand{\gammaL}{\textsc{Gamma}}
\newcommand{\http}{\textsc{HTTP}}
\newcommand{\linda}{\textsc{Linda}}
\newcommand{\sql}{\textsc{SQL}}
\newcommand{\xml}{\textsc{XML}}
\newcommand{\xpath}{\textsc{XPath}}
\newcommand{\yql}{\textsc{YQL}}

\newcommand{\bnfEq}{\ ::=& \ }

\newcommand{\lKind}[1]{& \text{{#1} } & }

\newcommand{\rKind}[1]{&  & \text{#1}}

\newcommand{\rKindTMTMTM}[6]{&  & \text{#1}#2\text{#3}#4\text{#5}#6}

\newcommand{\appSubst}[2]{{#1}[#2]}
\newcommand{\atome}{a}

\newcommand{\config}{\gamma}

\newcommand{\decPred}{D}

\newcommand{\elimDoublonsOp}{\Rrightarrow}

\newcommand{\equivautOp}{\equiv}
\newcommand{\nonEquivautOp}{\not\equiv}

\newcommand{\nonEquivaut}[2]{#1 \nonEquivautOp #2}

\newcommand{\etat}[2]{#1 \vdash #2}
\newcommand{\etJ}{\mathop{\&}} 
\newcommand{\etME}{\mathop{,}} 

\newcommand{\flot}[1]{{#1}^\omega}

\newcommand{\fv}[1]{\mathrm{FV(}#1\mathrm{)}}

\newcommand{\genere}[2]{{#1} \ \triangleright\  {#2}}

\newcommand{\join}{j}

\newcommand{\letIn}[2]{\texttt{let }{#1}\texttt{ in }{#2}}
\newcommand{\liste}[1]{\overrightarrow{#1}}

\newcommand{\multiRel}{M}

\newcommand{\para}{\mathop{,}} 
\newcommand{\predicat}{X}
\newcommand{\predicatY}{Y}
\newcommand{\porteeOp}{\nu}
\newcommand{\portee}[1]{\porteeOp\,{#1}.}
\newcommand{\progresseOp}{\Rightarrow}
\newcommand{\progresse}[2]{#1 \progresseOp  #2}

\newcommand{\processus}{p}

\newcommand{\reaction}{\rho}
\newcommand{\reduitOp}{\rightarrow}
\newcommand{\reduit}[2]{#1 \reduitOp #2}
\newcommand{\regle}{r}
\newcommand{\rel}{R}

\newcommand{\script}{s}
\newcommand{\seq}{\mathop{;}}
\newcommand{\soupe}{\sigma}
\newcommand{\subst}{\tau}

\newcommand{\var}{v}

\newcommand{\atom}{a}

\title{\creole{}:  a Universal Language for Creating, Requesting, Updating and Deleting Resources
\footnote{This work has been partially supported by the CESSA project (http://cessa.gforge.inria.fr/doku.php).}}

\author{Mayleen Lacouture
\institute{Ecole des Mines de Nantes\\ France}
\email{mayleen.lacouture@mines-nantes.fr}
\and
Herv\'e Grall
\institute{Ecole des Mines de Nantes\\
France}
\email{\quad herve.grall@mines-nantes.fr}
\and Thomas Ledoux
\institute{INRIA Rennes-Bretagne Atlantique \\ France}
\email{\quad thomas.ledoux@inria.fr}
}
\def\titlerunning{A CRUD language for Resource Manipulation}
\def\authorrunning{M. Lacouture, H. Grall \& T. Ledoux}

\begin{document}
\maketitle

\begin{abstract}
In the context of Service-Oriented Computing,
applications can be developed following the REST (Representation State Transfer) architectural style.
This style corresponds to a resource-oriented model,
where resources are manipulated via CRUD (Create, Request, Update, Delete) interfaces. 
The diversity of CRUD languages
due to the absence of a standard 
leads to composition problems related to
adaptation, integration and coordination of services.
To overcome these problems, we propose a pivot architecture
built around a universal language to manipulate resources, 
called \creole{}, a CRUD Language for Resource Edition.
In this architecture, scripts written in existing CRUD languages, 
like SQL,
are compiled into \creole{}
and then executed over different CRUD interfaces.
After stating the requirements for a universal language for manipulating resources, 
we formally describe the language and informally motivate its definition with respect to the requirements.
We then concretely show how the architecture solves 
adaptation, integration and coordination problems 
in the case of photo management in Flickr and Picasa, 
two well-known service-oriented applications.
Finally, we propose a roadmap for future work.
\end{abstract}

\section{Introduction}

The growth of Internet has extended the scope of software applications, leading
to Service-Oriented Computing (SOC):
it is a new computing paradigm that utilizes services as the basic
construct to develop distributed
applications, even in heterogeneous environments.
To date, there are two popular -- and often antagonistic -- models for
service-oriented computing~\cite{LeymannPautassoZimmermann2008}, which
we now describe as a process-oriented model and a resource-oriented
one. 

First, interoperability and integration issues have led to the
development of WS-* services technology, mainly based on XML and
SOAP.  
Upon services, which group together operations,
processes are defined with orchestration languages,
like the Business Process Execution Language for Web Services (BPEL),
which is a standard. 
As processes are central in this model, we say that this model is
process-oriented. 

More recently, 
an alternative solution has emerged thanks to
its simplicity:
RESTful Web services
return to the original design principles of the World Wide Web,
and its REST style~\cite{Fielding00}.
In this model, 
information and computation are  abstracted as \emph{resources},
which are manipulated using a fixed set
of four CRUD (create, read, update, delete) operations.
Since resources are central in this model, 
we say that the model is resource-oriented.
In a context analogous to databases,
CRUD languages for RESTful Web services have been developed as variants of the \sql{}
language: see for instance the language \yql{} from Yahoo.
But,
contrary to the process-oriented model, there is no standard like
BPEL, which has led to the current diversity of CRUD languages in use.

Because of the absence not only
of a unified model for service-oriented computing, 
but also of a standard for CRUD languages,
there is no 
universal language for  manipulating both services and resources,
which leads to some major issues, namely  
\emph{adaptation}, \emph{integration} and \emph{coordination} problems.
Let us illustrate these problems with two well-known Web photos
management systems, Picasa and Flickr.
Both  
provide CRUD interfaces  for client applications. 
However, their 
resource models and CRUD interfaces differ. 
Hence, an \emph{adaptation} is needed
when a client application that communicates with 
Picasa
must change to communicate instead with 
Flickr.
An \emph{integration} is needed when the client application
must communicate with both Picasa and Flickr. 
A \emph{coordination} is needed when two scripts, possibly written in
distinct languages, 
must cooperate to manipulate resources managed by one service. 

In this paper, we solve these problems in the simplest model, the
resource-oriented one. 
We propose a 
pivot architecture
built around a universal language for  manipulating resources. 
The pivot architecture decreases the coupling between CRUD
languages and CRUD interfaces, leading to a solution to the three
problems mentioned above for the resource-oriented model.  
Central to the pivot architecture, the  pivot language called \creole{} 
provides a
universal, minimalist and formal way of defining CRUD
scripts to manipulate resources.  

The paper is organized as follows.
First, after defining the problems of adaptation, integration and
coordination, 
we introduce the pivot architecture and present related work.
Second, we state  the requirements for a universal language for
manipulating resources and motivate its design, with respect to the
state of the art.  
Then, we describe the language \creole{}, its syntax and its semantics,
and validate its design against the requirements. 
Finally, we concretely show how the pivot architecture solves 
adaptation, integration and coordination problems 
in a paradigmatic use case, the management of photos in Flickr and Picasa.
We conclude by a roadmap for future work.  
An important step is to extend our solution, to deal not only with
the resource-oriented model, but also with 
the process-oriented model.

\section{A pivot architecture}
\label{sect-pivotArchi}

\begin{figure}[h!]
\centering
\includegraphics[width=.6\textwidth]{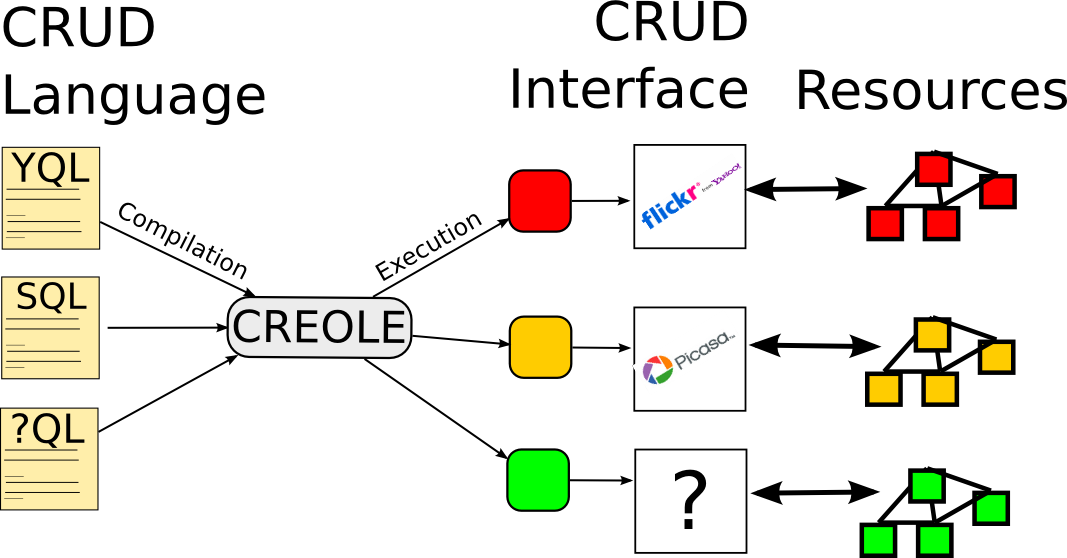}
\caption{A Pivot Architecture}\label{fig:papillon}
\end{figure}

The absence of a unified service-oriented language 
for manipulating CRUD resources leads to several interoperability
issues. Interoperability can be defined as the ability of two or more
systems or components to exchange information and to use the information
that has been exchanged\footnote{According to 
the IEEE Standard Computer Dictionary.}.
Without interoperability, we are faced with composition problems related
to adaptation, integration and coordination of heterogeneous services.  
By \emph{adaptation}, we mean the problem of switching from one service
provider to another without affecting its clients. 
By \emph{integration}, we mean the problem of providing a unified
interface for a set of resources managed by different CRUD interfaces. 
By \emph{coordination}, we mean 
the problem of executing different scripts,
possibly written in different languages, 
attempting to manipulate the same resources managed by one CRUD interface. 

The  pivot architecture described in Figure~\ref{fig:papillon} solves
these problems.
It is built around a universal language for  manipulating resources, called
\creole{} (CRUD Language for Resource Edition). 
Scripts written in existing CRUD languages, like \sql{}, are compiled into
the pivot language \creole{} and then executed over different CRUD
interfaces,  
like Picasa's or Flickr's.
To be effective, a pivot architecture relies on two assumptions.
First, it must be possible to compile from any 
source language to the pivot language. 
We will briefly see that the language \creole{} satisfies
this universality property with respect to CRUD languages. 
Thanks to this property, scripts written in
different CRUD languages can be coordinated by using the Mediator design
pattern~\cite[p.~273]{GoF}.  
Second, it must be possible to interface the language \creole{} 
with the applications manipulating resources, 
characterized by their own 
resource representation and CRUD interface. 
We will see that  
the design of these interface connectors,
called in the following built-in virtual
machines,
is akin to the design of 
RESTful Web services~\cite{LeymannPautassoZimmermann2008}.
We also use 
other virtual machines, dedicated to the execution
of the scripts written in the
pivot language \creole{}. 
To resolve
the adaptation and integration issues,
the virtual machines are organized following
two other design patterns,
namely the Adapter and the Facade patterns~\cite[pp.~139, 185]{GoF}, 
for adaptation and integration respectively. 

We identify several advantages of the pivot architecture.
First, using a pivot language avoids the combinatorial explosion 
of translations, from multiple CRUD languages to different CRUD interfaces.
Then, developers are allowed to program in their favorite CRUD language such as \sql{} or XQuery,
with the additional advantage of being able to profit from the specific features offered by each language.
Moreover, existing scripts written in different high-level languages can
be executed on different CRUD interfaces without the need to be
rewritten. 
%
%
%
Finally, the proposed pivot architecture overcomes
the composition problems related to adaptation, integration and
coordination.

\paragraph*{Related work}

In linguistics, a pivot language is an artificial or natural language
used as an intermediary language for easing translation between many different
languages (e.g. Interlingua, english).
In computing, 
for analogous reasons, pivot infrastructures built around an
intermediate language have been successful.
For instance, virtual machines with their bytecode language are now
common, allowing programs written in different languages to be compiled
and executed over different architectures and systems (e.g. Java VM, .NET). 

The pivot architecture can also benefit from techniques for the
generation of mediators, adapters and facades.
Instead of a manual generation as in
Section~\ref{sect:useCase-FlickrPicasa}, 
an automatic generation is possible, as exemplified
by Brogi and Popescu~\cite{BrogiPopescu2006} for \bpel{} processes,
and by Mateescu, Poizat and Sala\"{u}n for processes represented as symbolic
transitions systems and also implemented in \bpel{}~\cite{MateescuPoizatSalaun2008}. 
 
The main difficulty in a pivot architecture remains the design of the pivot
language and its associated virtual machine.
Various calculi, 
described in Bruni's comprehensive synthesis~\cite{Bruni2009},
have been proposed with the aim to capture 
aspects of service-oriented computing, 
from a verification or a modeling 
point of view but also from a formalization and programming point of
view, which is related to our approach. However, these calculi are
essentially process-oriented and not resource-oriented. 
As for the resource-oriented model, 
limited research have been undertaken in the formalization of RESTful
Web services.
Recently, Garrote and Moreno have proposed a 
language~\cite{GarroteMoreno2010} combining a  process
calculus for the exchanges of messages 
and the coordination language \linda{}~\cite{Gelernter1985}
for the description of resource computations.
Our solution presents the same two layers: a process language for
distribution and a script language for resource computations, which
as in Linda, includes operations for  
adding and  deleting data in a shared dataspace, as we will see in the
next sections.

%

\section{Requirements and design rationale for the pivot language}
\label{sect:requirementsRationale}

First, we attempt to identify some essential 
requirements for the language \creole{},
considered as the language for editing resources
at the heart of the pivot architecture. 
Second, we motivate the design with respect to the requirements.

\paragraph*{Requirements}

The requirements can be split into two parts: general ones, 
relative to 
service-oriented computing, and particular ones, relative to the
resource-oriented model.

Starting from the 
analysis led by Caires, Seco and Vieira 
\cite[Sect.~2]{CairesSecoVieira2008},
and a general presentation of service-oriented
computing~\cite{HuhnsSingh2005},  
we have identified four general requirements: distribution, process
delegation, scope management, and dynamic service binding.
We do not deal with the requirements about distribution and process delegation, 
already well described by Caires et al.~\cite{CairesSecoVieira2008}, 
but we focus on scope management and dynamic binding.

A client and a server execute in different contexts:
entities used in the execution can be either local
or shared between the server and the
client.
More interestingly,
contexts dynamically evolve.
For instance, 
a client can create a new session identifier that it sends to the server
with its request.
In its reply, the server also  transmits the identifier that the
client must use in order to relate the reply to its request. 
Thus, 
name creation and name extrusion turn out to be two essential
requirements.  
Name extrusion naturally leads to dynamic service binding, when the name
represents a service, via its location. 
Dynamic binding is used for 
service discovery~\cite[Fig.~1]{HuhnsSingh2005}
and dynamic routing, for instance
in a well-known service interaction 
pattern~\cite{BarrosDumasHofstede2005}
called 
Request with referral.

We have also identified requirements for the language
\creole{}
that are particular to the resource-oriented model:
they deal with resource modeling and its consequences 
for a pivot language.

How to represent a resource?
In the database field, 
since Codd's work, the data model has been defined as a relational
model. 
Likewise, the markup language \xml{}, used for representing data in web
services, is founded on a relational model, as shown for instance by
Benedikt and Koch's formalization of the query language 
\xpath{}~\cite{BenediktKoch2008}.
We require that the language \creole{} adopts the relational model to represent resources,
therefore assuming a logical approach.
Following model theory,
we represent resources as a structure, consisting of a universe and an
interpretation over the universe of each relation in some signature, 
used to define the class of the
resources considered. 

Choosing the relational model results in two requirements for \creole{},
since a pivot language must satisfy two properties, universality and
ability to interface, as seen in Section~\ref{sect-pivotArchi}.

The relational model is equipped with natural operators,
leading to the 
relational algebra: selection, projection, Cartesian product, set union, 
set difference, and renaming.  
We therefore require that the language \creole{} can express all these
operations. More generally, we require that the language can express any
computable transformation between structures:
the language must be universal with respect to the relational
model. 
For instance, 
it must be able to express 
aggregation and recursion, 
two powerful features, found natively but separately in 
\sql{}\footnote{See~\cite{Gogolla1994} for a formalization of \sql{}'s semantics.}  
and \datalog{}\footnote{See~\cite{CeriGottlobTanca1989} for an
introduction to \datalog.}
respectively.  

In the relational model, 
a resource is represented 
as a relational structure.
It has a uniform interface, namely a CRUD interface. 
A resource can be
created or deleted by adding its complete representation to the structure
or removing it respectively. 
It can be requested by querying the content of the structure and
updated by modifying the structure.
We therefore require that the 
CRUD interfaces of the relational structures can be mapped to the CRUD
interfaces of the resources managed by the applications to which the 
 language \creole{} is connected.

\paragraph*{Design rationale}

Just as the requirements are split into two parts,
the language \creole{} is designed with two layers, one defining scripts
for resource manipulation  
and one defining processes for distribution.

Consistent with our  logical point of view for representing
resources, our script language is first inspired by 
\datalog{}~\cite{CeriGottlobTanca1989},
a query language for deductive databases, in other words for structures
in the relational model. 
However, \datalog{} has a major limitation: it cannot express the
deletion or the update of resources. Its semantics is essentially
monotone: the representation of resources always increases during
computations.     
Several disconnected lines of research have addressed this
problem,
for instance Zaniolo et al. have extended \datalog{} with a
notion of choice~\cite{GrecoZaniolo2001} or with aggregate
operators~\cite{WangZaniolo2000}, 
and
Ganzinger and McAllester~\cite{GanzingerMcAllester2002} 
have allowed facts to be deleted and
rules to be selected with priorities. 
Instead of using ad-hoc extensions, we choose to use linear logic as a
foundation for our language.
Two recent works have directly inspired our work.

First, Pfenning and Simmons have proposed a programming language in
linear logic~\cite{PfenningSimmons2008}. 
Besides persistent predicates, as found in \datalog{}, there are
ephemeral predicates, corresponding to linear resources.
The operational semantics alternates a monotone deduction 
that involves only persistent predicates and a commitment
corresponding to the firing of a rule 
consuming ephemeral atoms, 
which are propositions 
built from ephemeral predicates.  
Second, Betz, Raiser and Frühwirth have defined an extension based on
linear logic for the
language \emph{Constraint handling Rules}~\cite{Fruhwirth2008} (\chr),  
a declarative language based on multiset rewriting, originally designed
for writing constraint solvers and now employed as a general purpose
language. 
They introduce persistent and
ephemeral predicates~\cite{BetzFruhwirthRaiser2010} in order to  
ensure termination for so-called propagation rules, leading to a
language akin to the preceding one.

Instead of 
using the distinction between persistent and ephemeral predicates, we
use a distinction between relations and multi-relations.
Multi-relations are multi-sets:
an element in a multi-relation may have multiple occurrences.
Relations are sets:
an element in a relation has a unique occurrence. 
Exhaustive duplicate eliminations 
transform a multi-relation into a relation. 
This distinction leads to a more primitive mechanism.
Indeed, 
whereas an ephemeral predicate is simply encoded as a multi-relation,
a persistent predicate is  encoded as  
a relation, and not a multi-relation, that satisfies an extra condition:
all atoms built
from a persistent predicate  must be 
preserved by rules. 
Persistence can therefore  be encoded.

Finally, 
generalizing 
the preceding languages
based on linear logic,
our script language is based on multiset rewriting.
Thus, it has also its
roots in the chemical reaction model: 
it can be considered as a variant of the 
language
\gammaL{}~\cite{BanatreLeMetayer1990}. 
More precisely, 
it is a restriction of a
coordination language with schedulers~\cite{Chaudron1996} for
a variant of \gammaL{}.
Indeed, we have considered as linear resources not only the atoms
but also the rules: rules are consumed when they are fired, except when
they are replicable. There is also a sequence operator, allowing rules
to be organized in distinct phases.

We now come to the distribution layer. Our process language is 
directly inspired by the join-calculus, a process calculus that can
also be considered as a language for multiset rewriting, with a chemical
semantics~\cite{FournetGonthier1996}. 
The join-calculus is interesting because of its natural notion of
location and its implementability in a
distributed setting. 
Rules are organized in definitions that are located. Given a channel,
which is equivalent to our notion of predicate (multi-relation or
relation),  
all the rules
consuming atoms built from this channel belong to the same definition.  
Whenever an atom is generated, it is migrated to the unique definition
dealing with the associated channel: this mechanism mimics a call from a
client to the
definition acting as a server. 
The join-calculus is also interesting because of its ability to express
dynamic binding: indeed, channels can be communicated.
Likewise, predicates can be communicated in \creole{}.


\section{Design and validation of the pivot language \creole{}}
\label{sect:creole}

We describe \creole{} syntax 
and its two layers, defining scripts for resource
manipulation  
and processes for distribution, respectively.
Table~\ref{tab-langageCreole} 
sums up this syntax\footnote{
As usual, we denote by $\fv{t}$ the set of free variables occurring
in the term $t$. The notation $\liste{x}$ denotes a sequence of $x$,
when the particular members of the sequence do not matter;
the sequence 
may be empty.}.
We then give the semantics of the language. 
We end the section by validating the design against the
requirements. 

\medskip

\begin{table}[!th]
{\small
 \vspace{-2em}
 \begin{align*}
   \lKind{Process} \processus \bnfEq{} (\liste{\predicat}) \script 
     \mid
     \letIn{\processus}{\processus} 
     \mid \processus\para\processus 
     \rKind{Script \underline{or} Let server used in client \underline{or} Parallel} \\
   \lKind{Script} \script \bnfEq{}  \emptyset \mid \regle 
     \mid \script\para\script \mid \script\seq\script \mid \flot{\script}
     \rKind{Skip \underline{or} Rule \underline{or} Parallel \underline{or} Sequence \underline{or} Replication} \\
   \lKind{Rule} \regle  \bnfEq{} 
    \genere{\join_1}{\liste{\portee{\var}}\join_2} 
    \rKindTMTMTM{If }{\join_1}{ then }{\join_2}{ with new
      names }{\liste{\var}} \quad \big(\liste{\var} = \fv{\join_2} - \fv{\join_1}\big) \\
   \lKind{Molecule} \join \bnfEq{} \emptyset \mid \atome \mid \join\etJ\join
    \rKind{Conjunction of atoms} \\
    \lKind{Predicate} \predicat \bnfEq{} \rel  \mid \multiRel
   \rKind{Relation \underline{or} Multi-relation} \\
   \lKind{Atom} \atome  \bnfEq{} 
     \predicat\,(\liste{\predicat},\liste{\var}) 
     \rKind{Predicate applied to predicates and variables} \\
   \end{align*}
 \vspace{-3em}
\caption{Language \creole{} -- Scripts and processes} 
 \label{tab-langageCreole} 
}
\end{table} 

\paragraph*{Scripts and processes}

The most primitive entities in \creole{} are 
predicates, either multi-relations or relations,
and variables. 
%
Atoms are built using
predicates and variables: 
a predicate $\predicat$ 
can be applied 
to a sequence $\liste{\predicatY}$ of predicates, possibly empty,
and to a sequence $\liste{\var}$ of variables, 
giving  atom
$\predicat(\liste{\predicatY}, \liste{\var})$. 
Atoms $\atom_1,\ldots,\atom_p$ can be joined together to make a molecule
$\atom_1\etJ\ldots \etJ\atom_p$.
The core part of \creole{} scripts  are reactions that transform
molecules into other molecules.
A reaction is specified by a rule
$\genere{\join_1}{\liste{\portee\var}\join_2}$, 
transforming any molecule matching the molecule pattern $\join_1$ to a
new molecule matching the molecule pattern $\join_2$, using new
variables in $\liste{\var}$. 
A variable in $\join_2$ is free if it occurs in $\join_2$  
without being declared in
$\liste{\var}$.  In that case, it must be bound by the rule: 
it must occur in $\join_1$.
Finally, a \creole{} script can be 
seen as a specification of a schedule for rules.  
There are basic scripts, the empty one, which contains no rule and does
nothing, 
and the singleton one, which contains a unique rule that can be fired
only once.  
The parallel operator allows scripts to be concurrently active.
For instance, 
the script $\regle\para\regle$ allows the rule $\regle$ to be fired
twice, whereas the script $\regle\para\regle'$ allows the rules $\regle$ and
$\regle'$ to be fired exactly once each one, in any order.
If a script needs to be executed an indefinite number of times, the
replication operator can be used: 
for instance, the script $\flot{\regle}$ means that the rule $\regle$
is always ready to be fired. 
There is also a sequential operator, at any depth, 
allowing the transformations
defined by scripts to be sequentially composed.

The distribution layer 
is defined around a process language:
a process distributes scripts in a client-server architecture.
The definition of a script is preceded with the declaration of 
the public predicates provided by the script.
Two processes can be put in parallel: they execute concurrently without
directly communicating.  
To enable a direct communication between two processes, 
the initial emitter or caller needs
to be declared as a client, 
and the initial receiver or callee as a server.
Consider the process $\letIn{\processus_{\mathrm{s}}}{(\decPred) \script}$, 
where $\processus_{\mathrm{s}}$ is the server process and
$(\decPred) \script$ the client process, equal to a simple script
$\script$ declaring public predicates in $\decPred$.
Each rule $\genere{\join_1}{\liste{\portee{\var}}\join_2}$
defined in script $\script$
can use in $\join_1$ and $\join_2$ 
the predicates in $\decPred$.
But it can also produce in $\join_2$ atoms built from 
predicates declared as public in the  server process
$\processus_{\mathrm{s}}$.
In other
words, a client can invoke a server.
How does the server reply to the client?
The client cannot directly consume atoms from predicates  
declared as public in the server process.
Indeed, this interaction  would violate the locality
principle that we impose to the process language, 
in conformity with the join-calculus~\cite{FournetGonthier1996}:
for each public predicate, there is one, and only one, script consuming this
predicate, the script where the predicate is declared, 
which allows a very simple implementation 
for atom communication. 
Actually, the client must transmit to the server a reference to one of
its own public predicate, which then can be used by the server to reply.
Thus, we introduce second-order predicates, 
$\predicat\,(\liste{\predicatY}, \liste{\var})$,
which are applied to
predicates $\liste{\predicatY}$ and 
variables $\liste{\var}$, 
in addition to first-order
predicates, 
$\predicat\,(\liste{\var})$,
only applied to variables $\liste{\var}$.
Thus, the rule $\genere{\join_1}{\liste{\portee{\var}}\join_2}$
can also produce in $\join_2$ atoms built from 
predicates  bound by $\join_1$.
A predicate used in the script that is neither public nor bound is
private: it is not usable outside of the script.

\paragraph*{Semantics with distributed chemical abstract machines}

The operational semantics of our script language is given by a reflexive
chemical abstract machine~\cite{FournetGonthier1996}. 
Due to the lack of space, it is informally given in this paper, with some
approximations.  
Its complete and accurate definition can be found 
in a technical report~\cite{GrallTabareau2010}.

A configuration $\config$ 
of the machine consists of two parts, $\etat{\reaction}{\soupe}$,
where $\reaction$ is the reaction part, 
a multiset of executing scripts,
and
$\soupe$ is the  solution part, a multiset of molecules.
There is a standard structural congruence between configurations,
as described by Berry and Boudol 
for the $\pi$-calculus~\cite{BerryBoudol1992}.  
It expresses for the multiset union -- denoted by a comma --
associativity, commutativity and 
neutrality of the empty script and of the empty
molecule -- both denoted by $\emptyset$ --,
and for the scope operator $\porteeOp$, 
the standard rules for name creation and
extrusion.
There are also two rules defining operators of the language:
{\small
\begin{align*}
 \textrm{Fusion and fission} & \quad & 
 \etat{\reaction}{\soupe\etME\join_1\etJ\join_2} &\equivautOp 
 \etat{\reaction}{\soupe\etME\join_1\etME\join_2}
 & \quad \quad & 
 \mathrm{Replication} & \quad & 
 \etat{\reaction\etME\flot{\script}}{\soupe} 
   &\equivautOp 
 \etat{\reaction\etME\flot{\script}\etME\script}{\soupe}
\end{align*}
}
Fission builds molecules from atoms whereas fusion is the reverse
operation, which gives the meaning of the join operator $\etJ$. 
As for the replication law, it gives the meaning of the
replication operator: a replicated script is always available for
execution. 

The execution of a configuration is defined in three steps. 
First the \emph{duplicate elimination} $\elimDoublonsOp$
eliminates every duplicated 
relational atom.
{\small
\begin{align*}
 \lKind{Duplicate elimination}
\etat{\reaction}{\soupe\etME\rel\,(\liste{\var}) \etME
  \rel\,(\liste{\var})} 
 &\elimDoublonsOp 
 \etat{\reaction}{\soupe\etME\rel\,(\liste{\var})}
\\
\end{align*}
}
The duplicate elimination, with possible fusions to decompose molecules,
is exhaustively performed between each reduction step to ensure that relational atoms
occur at most once in a configuration.
The \emph{chemical reduction} $\reduitOp$ describes the
basic reduction of the chemical abstract machine.
There are two main rules. 
{\small
\begin{align*}
 \lKind{Reaction}  
 \begin{prooftree}
  \justifies 
  \reduit{\etat{ \reaction\etME (\genere{\join_1}{
                         \liste{\portee{\var}}\join_2}) }{
                 \soupe\etME\appSubst{\join_1}{\subst} } }{ 
          \etat{\reaction
          }{    \soupe\etME
                \appSubst{(\liste{\portee{\var}}\join_2)}{\subst}
               } }
  \thickness=0.08em \using 
\end{prooftree}
\end{align*}
}
The first rule deals with the main mechanism, reaction. 
The reaction rule 
$\genere{\join_1}{\liste{\portee{\var}}\join_2}$ 
is fireable when a molecule matches the molecule pattern 
$\join_1$.
The firing  
generates 
a new molecule matching the molecule pattern $\join_2$, using new
variables in $\liste{\var}$; 
it consumes not only the molecule 
 matching the molecule pattern $\join_1$ but also the reaction rule.
{\small
\begin{align*}
 \lKind{Sequence} 
\begin{prooftree}
  \neg(\progresse{ \etat{\reaction_1}{\soupe} }{\ }) 
  \justifies 
  \reduit{ \etat{\reaction\etME(\reaction_1\seq\reaction_2)}{\soupe} }{
           \etat{\reaction\etME\reaction_2}{\soupe} }
  \thickness=0.08em \using 
\end{prooftree}
\end{align*}
}
The second rule deals with the sequence operator.
When the left part $\reaction_1$ of the sequence script does not
progress, it can be skipped. 
It remains to define  \emph{progression} $\progresseOp$ from reduction.
Assume that 
configuration $\config_1$ reduces to 
configuration $\config_2$:
$\reduit{\config_1}{\config_2}$.
After an exhaustive duplicate elimination,
configuration $\config_2$ becomes configuration $\config_3$.
We say that the machine progresses from $\config_1$ to $\config_3$, 
denoted $\progresse{\config_1}{\config_3}$, if
$\config_1$ is not structurally equivalent to $\config_3$,
$\nonEquivaut{\config_1}{\config_3}$. 
It means that either the reaction part, the solution part, or both, have
changed. 

Finally, the machine proceeds as follows.
Starting from an initial configuration with no duplicates, 
it looks for a progression, possibly by using the fusion and fission
rules and the replication rule.
If no progression can happen, then the configuration is final.
Otherwise, 
it non-deterministically chooses a possible progression, 
executes the associated reduction and exhaustively eliminates
duplicates in the resulting configuration.

Now, we come to the semantics of the process language. Given a process,
we associate to each script $(\decPred^+) \script$
declared in the process a chemical abstract
machine, called a virtual machine, having as interface 
the public predicates declared in $\decPred^+$.
A distributed configuration $\delta$ contains two parts,
a multiset of atoms $\atome$ 
migrating between virtual machines and 
a set of local configurations  
$[\config_i]_{\decPred_i}$,
where for each 
virtual machine $i$ associated to the process,
$\config_i$ is its local configuration and 
$\decPred_i = \decPred_i^+ \cup \decPred_i^-$ the
declaration of its predicates, either public (in $\decPred_i^+$) 
or  private (in $\decPred_i^-$). 
The progression relation between distributed configurations is an
extension of the progression relation defined for an individual virtual
machine. 
{\small
\begin{align*}
 \lKind{Local} 
\begin{prooftree}
  \progresse{ \config_i }{ \config'_i } 
  \justifies 
  \progresse{ \delta, [\config_i]_{\decPred_i} }{
              \delta, [\config_i']_{\decPred_i} }
  \thickness=0.08em \using 
\end{prooftree}
\end{align*}
}
It also contains two rules for the migration of atoms.
{\small
\[
\begin{array}{llll}
 \mathrm{Out} \quad &
\begin{prooftree}
  a = \predicat\,(\liste{\predicatY},\liste{\var}) \quad   
  \predicat \notin \decPred_i
  \justifies 
  \progresse{ \delta, [ \etat{\reaction_i}{\soupe_i, \atome} ]_{\decPred_i} }{
              \delta, [ \etat{\reaction_i}{\soupe_i} ]_{\decPred_i} , \atome }
  \thickness=0.08em \using 
\end{prooftree}
 &
 \quad \mathrm{In} \quad
 &
\begin{prooftree}
  a = \predicat\,(\liste{\predicatY},\liste{\var}) \quad   
  \predicat \in \decPred_i^+
  \justifies 
  \progresse{ \delta, [ \etat{\reaction_i}{\soupe_i} ]_{\decPred_i} , \atome}{
              \delta, [ \etat{\reaction_i}{\soupe_i, \atome} ]_{\decPred_i} }
  \thickness=0.08em \using 
\end{prooftree} 
\end{array}
\]
}

After we have defined the syntax and the semantics of the language
\creole{}, we now assess the design with respect to the requirements
presented in Section~\ref{sect:requirementsRationale}. 

\paragraph*{Validation against requirements} 

For validation, we consider the following requirements:
distribution, with two aspects -- implementation and expressivity --,
scope management and dynamic service binding,
script expressivity and ability to interface.  

Thanks to its distributed semantics, implementing the language in a
distributed context is easy. It suffices to assign to each script and
its associated virtual machine a definite location like a 
Uniform Resource Locator (URL). Then each atom built from a public
predicate needs to convey the location where the predicate is declared, 
in order to 
allow the atom to be migrated when it is produced in another virtual
machine.  
The only communication primitive in \creole{} is 
atom migration,
corresponding to an asynchronous one-way invocation. 
As an atom can contain as argument a predicate for
reply, dynamic binding is present for predicates,
allowing  different 
request-reply 
interactions, synchronous or not, to be encoded.
For instance, let $s$ be the following server script: 
$\flot{(\genere{I(K)}{K()})}$. Then an echo interaction can be
described as follows:
$$
\letIn{(I)s}{
 (K)\big(
  (\genere{\emptyset}{I(K)}),(\genere{K()}{\ldots})
 \big)}.
$$

Scope is statically managed for predicates, 
with the distinction between public and
private predicates. 
Name creation is available for variables, 
and name extrusion for predicates and
variables. 
Thus, a virtual machine can control its state and share relevant names. 
For instance, the precedent example can be refined in order to manage a
session, allowing the reply to be related to the request:
$$
\letIn{(I) \flot{(\genere{I(x, K)}{K(x)})} }{
 (K)\big(
  (\genere{\emptyset}{\portee{x} I(x, K) \etJ W(x)}),
  (\genere{W(x) \etJ K(x)}{\ldots})
 \big)}.
$$

As for the expressivity requirements with respect to the relational
model, it is easy to show that any operation in the relational algebra
can be encoded in our script language.   
Aggregation can also be encoded.
For instance, the script 
$$
\genere{\emptyset}{C(0)}, 
 \flot{\big( \genere{C(n) \etJ R(x)}{C(n+1)} \big)}
$$ 
counts the number of elements in predicate $R$, assuming the availability
of natural numbers, which can also be  encoded in a straightforward
manner. 
It is therefore possible to encode any \sql{} query in our script
language, allowing the definition of a compiler. 
As for recursion, it is natively supported by our script language. 
For instance, \datalog{} with negation, equipped with its well-founded
semantics, can be encoded~\cite{GrallTabareau2010}.

In the relational model,
resources are represented 
as relational structures, using multi-relations when the number of
occurrences matters, and using relations otherwise. 
All CRUD operations over relational structures  can be mapped to \http{}
operations over resource representations,
precisely to
\texttt{PUT}, 
\texttt{GET},
\texttt{POST} 
and \texttt{DELETE}
respectively. 
This correspondence paves the way for an implementation with RESTful Web
services of the built-in virtual machines, 
connecting the language \creole{} 
to the applications manipulating resources.

Thus, the language \creole{} satisfies the requirements that we have
defined. 
In the next section, 
we illustrate the use of our pivot architecture and language in a
paradigmatic use case.

\section{Use Case: Photo Management on Flickr and Picasa}
\label{sect:useCase-FlickrPicasa}
 
Flickr and Picasa are  Yahoo's and Google's respective photo management systems.
They offer web interfaces (APIs) 
to enable  client applications to publish and organize photos on-line.
These interfaces are essentially CRUD interfaces, 
implemented as RESTful web services and 
allowing photos to be
manipulated as resources. 
This section illustrates the use of our pivot architecture and of
\creole{} to solve adaptation, integration and coordination problems 
in the case of photo management with Flickr and Picasa.
Concretely, our solution is based on  three general
design patterns, Adapter, Facade 
and Mediator.

\paragraph*{Problem I: Adaptation}

Yahoo proposes a \sql{}-like language called \yql{} to query 
web services as if they were tables.
In \yql{}, web services are represented as \emph{virtual tables} 
wherein columns are mapped to input and output parameters.
For example, the Flickr CRUD interface 
contains a method called \texttt{flickr.photo.counts}
to count photos in a given date range.
This service is represented in \yql{} as a virtual table called \texttt{PhotoCounts}
with \texttt{fromDate} and \texttt{toDate} as  input columns and 
\texttt{count} as an output column.
The method can be called from a  
\yql{} query, akin to a \sql{} query:

\begin{small}
\[
\begin{array}{l}
\texttt{SELECT count FROM PhotoCounts} \\
\texttt{WHERE fromDate ="01/01/2009" AND toDate="31/12/2009"}
\end{array}
\]
\end{small}
This script is mapped to a call to  method \texttt{flickr.photo.counts}.
Columns \texttt{fromDate} and \texttt{toDate} correspond to the method's input parameters,
and column \texttt{count} to one of the output parameters.

\begin{figure}[ht!]
\centering
(a)\includegraphics[width=.4\textwidth]{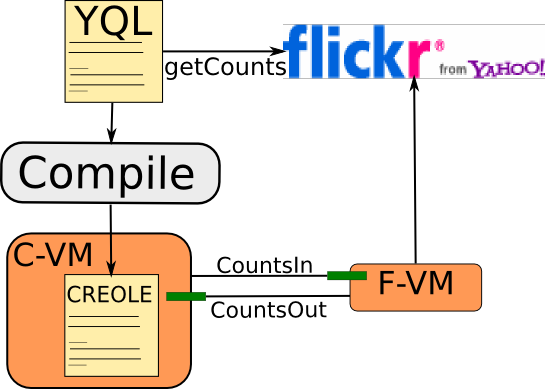}(b)\includegraphics[width=.49\textwidth]{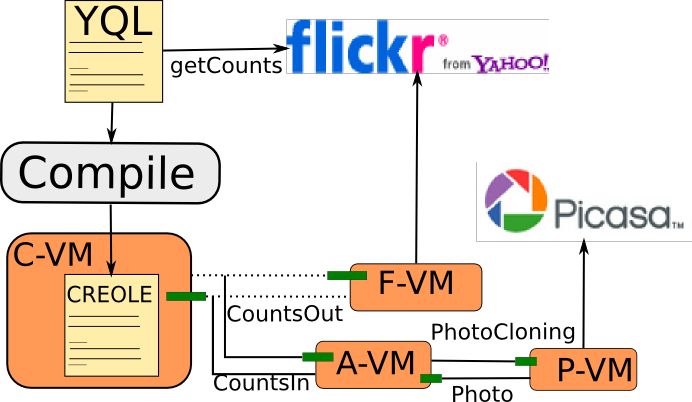}
\caption{Flickr-Picasa Adaptation}\label{fig:adaptation}
\end{figure}

How can we adapt the \yql{} script to count photos on Picasa,
knowing that its CRUD Interface does not offer a count  operation?
Figure~\ref{fig:adaptation} summarizes our approach.
In (a), the \yql{} script is compiled into \creole{}, then executed on a
virtual machine (C-VM). 
In (b), we implement a virtual machine (A-VM), an Adapter
allowing to switch from Flickr to Picasa.  
In this schema, virtual machines can be compared to components 
whose provided interfaces are relations,
represented here by flat rectangles.
We now detail the approach.

To compile the \yql{} script into \creole{},
we map the virtual table \texttt{PhotoCounts} to relations 
\texttt{CountsIn} and \texttt{CountsOut}.
Input columns \texttt{fromDate} and \texttt{toDate} are mapped to
\texttt{CountsIn}'s parameters, 
and output column \texttt{count} is mapped to \texttt{CountsOut}'s last
parameter.  
The following is the resulting \creole{} script:
%
\begin{small}
\[
\begin{array}{ll}
\textrm{1:}& \genere{\emptyset}{\portee{x}\texttt{CountsIn}(x,"01/01/2009","31/12/2009", \texttt{CountsOut})\etJ \texttt{Session}(x)},\\
\textrm{2:}& \genere{\texttt{Session}(x) \etJ \texttt{CountsOut}(x,n)}{\texttt{Result}(n) }
\end{array}
\]
\end{small} 
%
The fresh variable $x$, representing a session identifier,  
is used to relate the reply to the request. Note that the request
transmits the relation \texttt{CountsOut} where it will obtain the reply. 

The compiled \yql{} script is executed on a client virtual machine (\texttt{C-VM}).
This virtual machine 
communicates with Flickr's built-in virtual machine (\texttt{F-VM})
which serves as a connector to Flickr's CRUD interface.
Built-in virtual machines, like \texttt{F-VM}, 
are programmed to map CRUD operations over relations
to RESTful Web services, accessed by \http{} requests. 

In the second part of our solution, we create an adaptation virtual machine (\texttt{A-VM})
to adapt the desired behavior to Picasa's built-in virtual machine (\texttt{P-VM}).
The following is the script executed in 
\texttt{A-VM}\footnote{For readability, the rule $\genere{R \etJ S}{R
    \etJ Q}$ is 
  written $\genere{\underline{R} \etJ S}{Q}$, where we have underlined
  the  persistent atom $R$. We also use natural numbers, and the
  relations  \texttt{Between} and  \texttt{NotBetween} to compare dates,
  and \texttt{Null} and  
  \texttt{NotNull} to test nullity,
  assuming their availability.}.
\begin{small}
\[
\begin{array}{ll}
\textrm{1: } \big( &  
 \genere{\texttt{CountsIn}(x,from,to,K)}{
         \portee{y} 
            \texttt{Response}(x, y, from, to, 0, K) 
           \etJ \texttt{PhotoCloning}(\texttt{Photo},y)}, \\
\textrm{2: } \ \ \big( & 
  \underline{\texttt{NotNull}}(id) 
  \etJ \underline{\texttt{Between}}(from,date,to) 
  \etJ \texttt{Response}(x, y,from, to, n, K) 
  \etJ \texttt{Photo}(y, id, date) 
  \ \triangleright \\
                       &  \quad 
  \texttt{PhotoCloning}(\texttt{Photo}, y) 
  \etJ \texttt{Response}(x, y, from, to, n+1, K), \\
\textrm{3: }         & 
  \underline{\texttt{NotNull}}(id) 
  \etJ \underline{\texttt{NotBetween}}(from,date,to) 
  \etJ \underline{\texttt{Response}}(x, y, from, to, n, K) 
  \etJ \texttt{Photo}(y, id, date) 
  \ \triangleright \\
                     & \quad 
  \texttt{PhotoCloning}(\texttt{Photo}, y)\quad \big) \flot{}, \\
\textrm{4: }         & 
  \genere{\underline{\texttt{Null}}(id) 
          \etJ \texttt{Photo}(y,id,date) 
          \etJ \texttt{Response}(x, y, from,to, n, K)}{
          K(x,n) 
  } \quad \big) \flot{} 
\end{array}
\]
\end{small} 
To count the photos taken between the two dates,
the script uses
 \texttt{P-VM}'s relation \texttt{PhotoCloning}.
Given an identifier $y$,
when the built-in virtual machine 
\texttt{P-VM} receives a request
\texttt{PhotoCloning}$(K, y)$ for the first time, it produces 
a relation containing the relevant photos,
by addressing \http{}  \texttt{GET} requests
to the Picasa server and  
answers by sending a first photo using the relation $K$.
Then at each request $\texttt{PhotoCloning}(K, y)$, 
\texttt{P-VM} sends a new photo of the relation produced over $K$.
When there is no more photo in the relation,
it sends a photo with null as identifier.
In the \texttt{A-VM} script, 
each time a photo is received, a request for another photo is sent
(cf.~lines~2~and~3);
moreover, when the date of the photo satisfies the comparison criterion, the
counter is incremented.
When the null identifier is received, indicating that there are no more
photos, the answer is sent to the client (cf.~line~4).
The whole script is replicated in order to indefinitely satisfy requests.

Finally, to switch from Flickr to Picasa,
all we need to do is to change the virtual machine used as a server,
from \texttt{F-VM} to \texttt{A-VM}, as follows:
\begin{small}\texttt{let (CountsIn, ...) A-VM in C-VM}\end{small}.


\paragraph*{Problem II: Integration}

Despite the fact that both Picasa and Flickr manage similar resources,
most of the time they are not represented in the same way.
For instance, if photos in Picasa are represented by the relation 
$\texttt{Photo}(id, date, \liste{x})$,
then in Flickr they are represented by the relation 
$\texttt{Photo}(id, date, \liste{y})$,
where $\liste{x}$ and $\liste{y}$ do not have the same elements.

\begin{figure}[th]
\centering
\includegraphics[width=.75\textwidth]{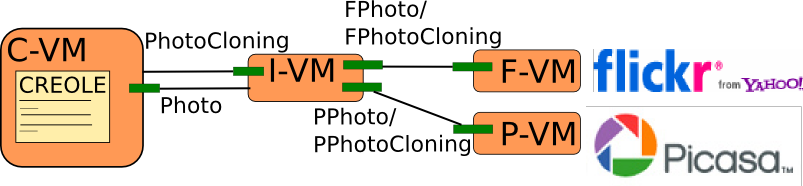}
\caption{Flickr-Picasa Integration}\label{fig:integration}
\end{figure}

Nevertheless, \creole{} facilitates the implementation of an integration solution,
like the one shown in Figure~\ref{fig:integration}.
In this schema, an intermediate virtual machine (\texttt{I-VM}),
implementing a Facade,
provides a common representation for photos in Flickr and Picasa,
which is then used by the client virtual machine (\texttt{C-VM}).

In this configuration,  
built-in virtual machines \texttt{F-VM} and \texttt{P-VM}
provide both a relation to obtain photo information.
As above, we call these relations 
\texttt{FPhotoCloning} and \texttt{PPhotoCloning}. 
Since the intermediate virtual machine (\texttt{I-VM})
holds a common representation for photos,
it also provides  a relation 
\texttt{PhotoCloning},
combining the attributes of Picasa's and Flickr's photos in the response.
The following is the script corresponding to  \texttt{I-VM}.
%
\begin{small}
\[
\begin{array}{ll}
\textrm{1: } \big( &  
  \genere{\texttt{PhotoClonning}(P, x)}{
          \texttt{PPhotoCloning}(\texttt{PPhoto}, x) 
          \etJ \texttt{Response}(P,x)}, \\
\textrm{2: } \ \ \big( & 
  \genere{\underline{\texttt{NotNull}}(id) 
          \etJ \texttt{PPhoto} (x, id, date, \liste{p}) 
          \etJ \underline{\texttt{Response}}(P,x) } {} \\
                       & \quad
          P(x, id, date, \liste{p'}) 
          \etJ \texttt{PPhotoCloning}(\texttt{PPhoto},x), \\
\textrm{3: }           & 
  \genere{\underline{\texttt{Null}}(id) 
          \etJ \texttt{PPhoto} (x, id, date, \liste{p}) } {
          \texttt{FPhotoCloning}(\texttt{FPhoto},x)}, \\
\textrm{4: }           & 
  \genere{\underline{\texttt{NotNull}}(id) 
          \etJ \texttt{FPhoto} (x, id, date, \liste{f}) 
          \etJ \underline{\texttt{Response}}(P,x) } {} \\
                       & \quad
          P(x, id, date, \liste{f'}) 
          \etJ \texttt{FPhotoCloning}(\texttt{FPhoto},x) 
          \quad \big) \flot{}, \\
\textrm{5: }           & 
  \genere{\underline{\texttt{Null}}(id) 
          \etJ \texttt{FPhoto} (x, id, date, \liste{f}) 
          \etJ \texttt{Response}(P,x)}{
          P(x,id,date,\liste{f'})} 
  \quad \big ) \flot{} 
\end{array}
\] 
\end{small} 
%
The lists $\liste{p'}$ and $\liste{f'}$ contain the same attributes
and are computed from  some combination, between intersection and
union, of 
attributes in $\liste{p}$ and $\liste{f}$ respectively.
As a consequence, we can simultaneously execute queries of photos on both CRUD interfaces.
For example, we could execute the script of the adaptation scenario
to count all our photos on both Flickr and Picasa,
by setting \texttt{I-VM} as the server instead of \texttt{P-VM}:
\begin{small}\texttt{let (PhotoCloning, ...) I-VM in A-VM}\end{small}.

Due to the lack of space, we have presented a simple scenario;
nevertheless, there are more complicated differences between Flickr and
Picasa  that can be tackled with our approach.
Consider, for instance, how photos are organized in both services:
in Flickr, photos can be organized in sets but can also be on their own;
in Picasa however, photos must belong to one and only one album.
We can solve this problem by using a common representation for albums and sets,
and then applying the same integration schema as in the example shown here.

\paragraph*{Problem III: Coordination}
\label{coordination}

\begin{figure}[th!]
\centering
\includegraphics[width=.75\textwidth]{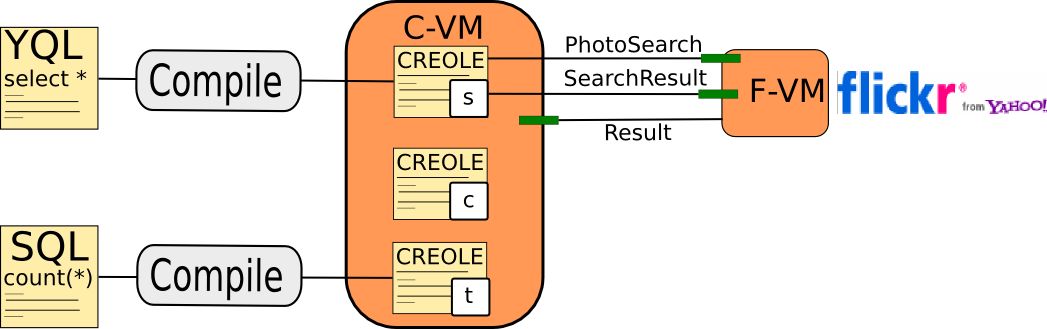}
\caption{Coordination}\label{fig:coordination}
\end{figure}

One of \yql{}'s limitations is the lack of support for aggregation.
With \creole{} it is possible to coordinate scripts written in different languages 
to take advantage of features provided by each language.
Hence,
we can combine \yql{} capacity for querying services as tables
with \sql{} support for aggregation. 
In the example shown in Figure~\ref{fig:coordination}, 
a \yql{} script to select photos taken between 01/01/2009 and 31/12/2009
is coordinated with a \sql{} script that counts rows from a given relation.
Here are the corresponding \yql{} and \sql{} queries:
\begin{center}
\begin{small}
\begin{tabular}{l | l}
 \texttt{SELECT * FROM PhotoSearch}  & \texttt{SELECT COUNT(*) FROM R} \\
 \texttt{WHERE min\_taken\_date="01/01/2009"} \\
 \texttt{AND max\_taken\_date="31/12/2009"} 
\end{tabular} 
\end{small} 
\end{center}
The virtual table \texttt{PhotoSearch} is a representation of Flickr's method \texttt{flickr.photo.search}  
which takes \texttt{min\_taken\_date} and \texttt{max\_taken\_date} as input parameters. 
Note in the \sql{} query that \texttt{R} can be any relation since the query is
not bound to a concrete database implementation. 

The \yql{} and \sql{} queries are compiled into \creole{} 
and executed on a coordination virtual machine (\texttt{C-VM}).
The \texttt{C-VM} virtual machine
uses the relations \texttt{PhotoSearch} 
and \texttt{SearchResult}
provided by the built-in virtual machine \texttt{F-VM}.
Given an identifier $x$,
when \texttt{F-VM} receives a request 
$$
 \texttt{PhotoSearch}(a, b,x),
$$
it produces 
a relation associated to $x$ and containing the photos taken between $a$
and $b$,
by addressing \http{}  \texttt{GET} requests
to the Flickr server.
Then, 
at each request $\texttt{SearchResult}(K, x)$, 
\texttt{P-VM} sends a new photo of the relation produced over $K$.
The following script $s$ is the \yql{} query compiled into \creole{}:
\begin{small}
\[
\begin{array}{ll}
\textrm{1: } &   
  \genere{\emptyset}{\portee{x} 
     \texttt{PhotoSearch}("01/01/2009", "31/12/2009", x) 
     \etJ \texttt{SearchResult}(\texttt{Result}, x)}, \\
\textrm{2: } &  \big( 
  \genere{\underline{\texttt{NotNull}}(id) 
          \etJ \texttt{Result}(x,id,\liste{y})}{
          \texttt{SearchResult}(\texttt{Result}, x) \etJ \texttt{Photo}(x,id,\liste{y})} 
  \big) \flot{},\\
\textrm{3: } &   
  \genere{\underline{\texttt{Null}}(id) 
          \etJ \texttt{Result}(x, id, \liste{y})}{
          \emptyset}
\end{array}
\] 
\end{small} 
The script $s$ initiates the search by calling the server \texttt{F-VM}.
Then, 
each time a photo is received, a request for another photo is sent
(cf.~line~2).
Finally, when there is no more photo, the script ends.
At the same time, the \sql{} query is compiled into the following script
$t$:
\begin{small}
\[
\genere{\emptyset}{\texttt{Count(0)}}, 
  \flot{\big( 
   \genere{\texttt{Count}(n) \etJ R(\liste{y})}{\texttt{Count}(n+1)} 
  \big)} 
\] 
\end{small} 
Finally, a third script $c$ coordinates the previous
scripts, implementing a Mediator: 
\begin{center}
\begin{small}
 $\flot{\big( \genere{\texttt{Photo}(\liste{y})} {\texttt{R}(\liste{y})} \big)}$ 
\end{small} 
\end{center}
This script, in parallel with $s$ and $t$, 
combines the outcomes of the \yql{} script $s$  and the counting of the 
\sql{} script $t$ with a renaming from \texttt{Photo} to \texttt{R}. 
It finally produce an atom $\texttt{Count}(n)$, where $n$ is the number
of photos.

\section{Conclusion and Future Work}

In the context of Service-Oriented Computing, we have identified three main problems related to service composition,
namely  \emph{adaptation}, \emph{integration} and \emph{coordination},
due to the absence of a unified model for manipulating resources.
We have presented our approach to tackle these problems,
consisting of a pivot architecture,
where existing languages for manipulating resources
are compiled into a pivot language, called \creole{}, 
and then executed over different resource interfaces, which are CRUD
interfaces. 
We have mainly introduced \creole{}, 
a universal language for resource manipulation,
which is at the heart of our solution.
The motivating example of photo management on services like Flickr and
Picasa
has concretely shown
how our proposed architecture solves  
adaptation, integration and coordination problems,
and how \creole{} can be used either 
as a CRUD language or as a target language
for the compilation from existing CRUD languages.

Yet we have only explored the resource-oriented model for services.  
An extension towards the process-oriented model would be valuable:
indeed, it will bring a unified foundation for service-oriented
computing.
Actually, the two models share a lot of similarity, since they follow a
same architecture with three layers.
First, there are resources. Second, there are services, limited to  
CRUD operations for the resource-oriented model and extended to any
computation for the process-oriented model.  Third, there are processes
or scripts 
for orchestrating services.

Our future work has therefore two main objectives.
First, we want to develop the formal foundations of the language
\creole{}, as begun in our technical report~\cite{GrallTabareau2010}.
The main questions here are the development of the theory of the
language, from operational semantics to bisimilarity,
and the assessment of its expressive power. 
Second, we want to implement the language and the whole pivot
architecture. 
Four questions are here important: implementation of the chemical
abstract machine and of its distribution, design of compilers 
into \creole{}
for existing 
languages like \yql{} and \bpel{}, 
design of a user-friendly 
programming language based on the core calculus 
presented here, 
implementation of built-in virtual machines 
by connecting to RESTful services and WS-* services.
Thus, these objectives pave the way to a unified foundation for
service-oriented computing, in a theoretical and practical perspective.

\paragraph{Acknowledgments}
We are grateful to the anonymous reviewers of FOCLASA 2010 for their useful suggestions to improve this paper.

\bibliographystyle{eptcs}
\bibliography{foclasa}

\end{document}